\documentclass[fleqn,twoside]{article}
\usepackage{espcrc2}
\usepackage[dvips]{graphics}

\begin{document}

\title{Masses of singlet and non-singlet 
$0^{++}$
particles.}

\author{UKQCD Collaboration:
Alistair~Hart\address{
DAMTP, CMS, University of Cambridge, Wilberforce Road, 
Cambridge CB3 0WA, UK},
Craig~McNeile\address[LIV]{Department of Mathematical Sciences, 
University of\ Liverpool, L69 3BX, UK}
\thanks{Poster presented by Craig McNeile}
and  Chris~Michael\addressmark[LIV].
}

\begin{abstract}
We compute the mass of the singlet $0^{++}$ state using
both $\overline{\psi}\psi$ and Wilson loop operators from a $N_f=2$
lattice QCD calculation.
\end{abstract}
 
\maketitle

\section{INTRODUCTION}

The $0^{++}$ singlet state is an excitation of the vacuum.
The complicated non-perturbative nature of the vacuum
is one of the reasons that QCD is hard to solve at low energies.
The precise computation of scalar meson masses via a first
principles full QCD calculation would demonstrate that 
low energy QCD could be tamed.

The work of many 
groups~\cite{Bali:1993fb,Morningstar:1999rf,Vaccarino:1999ku} has
shown that the lightest $0^{++}$ state is 1611(30)(160) 
MeV~\cite{Michael:2001qz}
in pure SU(3) Yang-Mills theory.
The priority now is to find evidence for glueballs in nature. 
This almost certainly requires an understanding of the mixing of glueballs
with quark anti-quark states.
Experimentally, there are more
singlet $0^{++}$ states in the region 1200 to 2000 MeV than can be
organised into SU(3) nonets~\cite{Close:2001zp}.  
An understanding of the non-singlet $0^{++}$ states is also
required to show there are additional degrees of freedom
beyond  $\overline{\psi}\psi$ states.

In full QCD all operators with $J^{PC}$ = $0^{++}$ quantum numbers can
mix. The singlet $\overline{\psi}\psi$ operator also has $0^{++}$
quantum numbers, hence these operators will mix with Wilson loop
$0^{++}$ operators used for glueballs in quenched QCD.  The mixing of
Wilson loop and $\overline{\psi} \psi$ states has been studied in
quenched QCD by Lee and Weingarten~\cite{Lee:1999kv}. They claimed
that in the continuum limit the mixing between Wilson loop and
$\overline{\psi}\psi$ $0^{++}$ states is small.  Constructive
criticism of the Lee-Weingarten calculation is 
in~\cite{Michael:2001qz,McNeile:2000xx}.

In this work we report results for the masses of 
singlet and non-singlet $0^{++}$ states.  We use 
unquenched gauge configurations~\cite{Allton:2001sk}
with a smaller lattice spacing ($a \sim $ 0.1 fm) 
than our previous study~\cite{McNeile:2000xx} 
($a \sim $ 0.13 fm). 

\section{METHOD}

To extract masses we use factorising (or variational) fits of $M$
exponentials to correlators of operators $\{ O_i \}$,
\begin{equation}
\langle O_i^{\dagger}(t) O_j(0) \rangle =
\sum_{n=0}^M c_i^n c_j^n e^{- E_n t} .
\label{eq:matrixFIT}
\end{equation}
Using additional operators helps to stabilise
the multi-exponential fit. We include $0^{++}$ operators made out of
a quark and anti-quark and Wilson loops in the same fit.
This basis of operators should couple well to the glueball mixing.

The Wick contraction of the singlet $\overline{\psi}\psi$ 
operators requires the calculation of fermion 
loops~\cite{McNeile:2000xx}.
\begin{equation}
\langle 
  \overline{\psi}(t,\underline{x}) \psi(t,\underline{x}) 
  \overline{\psi}(0,\underline{0}) \psi(0,\underline{0}) 
\rangle
\label{eq:singletMESON}
\end{equation}
The correlator in equation~\ref{eq:singletMESON} can be computed
using $Z_2$  noise techniques. We used 100 noise sources
and the double source techniques described in~\cite{McNeile:2000xx}.
In the fermion sector we use fuzzed~\cite{Allton:2001sk} and local
operators as basis states.  Two types of smeared Wilson
loop operators~\cite{Hart:2001fp} were included.
We use $M=1$ and 2 in this calculation.

The non-perturbatively improved
clover action was used to generate the configurations,
with $c_{SW} = 2.02$, $\beta = 5.2$, and 
volume of $16^3 32$~\cite{Allton:2001sk}.
The configurations with sea $\kappa = 0.135$
and $\kappa = 0.1355$ were used.
We use $f_0$ ($a_0$) to label the singlet 
(non-singlet) $0^{++}$ states.

\section{RESULTS FOR THE \protect{\mbox{$a_0$}}}

The interpretation of the experimental spectrum of the $a_0$ 
hadrons is confused by the $a_0(980)$, that many people
believe is a $\overline{K}K$ molecule~\cite{Close:2001zp}.
In quenched QCD the $a_0$ channel is complicated by a
severe quenched chiral artifact~\cite{Bardeen:2001jm}, caused
by the $\pi-\eta'$ state being treated incorrectly in
quenched QCD.

One prediction~\cite{Bardeen:2001jm} of the
analysis of Bardeen et al. is that the mass of the $a_0$
should not reduce smoothly with quark mass,
but actually start to increase. This behavior was observed by Lee and
Weingarten~\cite{Lee:1999kv}, who
 saw the mass of the $a_0$ rise with quark mass below
that of strange for small volumes (less than 1.8 fm).
Other groups have seen similar 
behavior~\cite{Gockeler:1998fn,Alford:2000mm}.

\begin{figure}[htb]
\vspace{-2.7cm}
\begin{center}
\scalebox{0.4}{\includegraphics{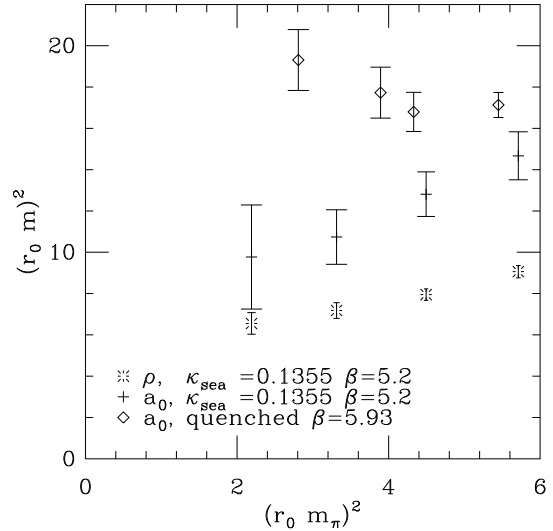}}
\end{center} 
\caption{
Mass dependence of the $a_0$ and $\rho$ in quenched
and partially quenched QCD.
}
\label{eq:a0MASS}
\end{figure}

The $a_0$ channel was fitted using a variational basis of 
local and fuzzed operators.
We do a partially quenched
analysis, where the sea quark mass is fixed and the valence
quark mass is varied. The partially quenched theory also
has pathologies that tend to be less severe than for the 
quenched theory. In figure~\ref{eq:a0MASS}, we plot the
mass of the $\rho$ and $a_0$ as a function of pseudoscalar
mass squared, using $r_0$~\cite{Sommer:1994ce} to 
make a dimensional ratio, in both 
quenched ($\beta=5.93$) and unquenched QCD.
Figure~\ref{eq:a0MASS} does show the quenched pathology predicted
by~\cite{Bardeen:2001jm}.
The results in figure~\ref{eq:a0MASS} convince us
that it is preferable  to study the $a_0$ mass in unquenched
lattice calculations.
If we chirally extrapolate the $a_0$ mass for the unquenched
data in figure~\ref{eq:a0MASS} to $\kappa_{critical}$ we get
$1.0(2)$ GeV. More work is required to control the systematic
errors on the mass of the $a_0$, so that lattice QCD calculations
can distinguish between the mass of the $a_0(980)$ and
$a_0(1450)$.

\section{RESULTS FOR THE \protect{$f_0$}}

The results from unquenched QCD lattice calculations, with light quark
masses and fine lattice spacings, should automatically include the
physics of glueball $\overline{\psi}\psi$ mixing.  However, the
results from $n_f=2$ lattice calculations from the
HEMCGC~\cite{Bitar:1991wr} and SESAM~\cite{Bali:2000vr} collaborations
did not show a large deviation of the masses of $0^{++}$ glueballs
between quenched and unquenched QCD.

\begin{figure}[htb]
\begin{center}
\scalebox{0.4}{\includegraphics{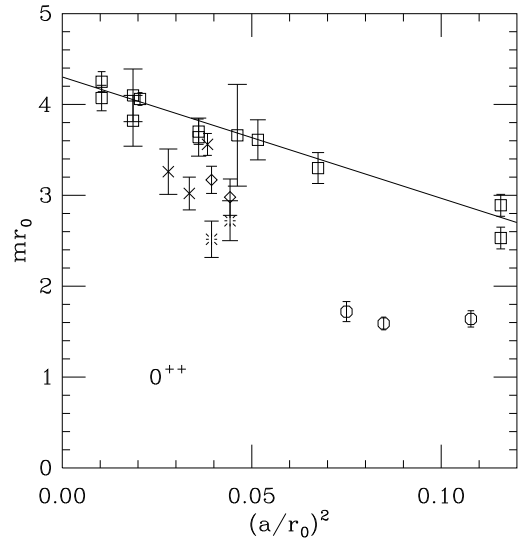}}
\end{center} 
\caption{\protect{$f_0$} mass in units of 
\protect{$r_0$}
as a function
of lattice spacing.
The crosses are from SESAM~\cite{Bali:2000vr}. 
The octagons are from UKQCD's~\cite{McNeile:2000xx}
first $n_f = 2$ data set. 
The diamonds are the results from Hart and Teper~\cite{Hart:2001fp}
and the bursts are from the analysis described in this paper.
The squares are the results from quenched 
calculations (see~\cite{McNeile:2000xx} for references).
}
\label{eq:GlueCont}
\end{figure}

In figure~\ref{eq:GlueCont} we show a compendium of lattice results
for the mass of the $f_0$ state in quenched and
$n_f$ = 2 QCD~\cite{McNeile:2000xx}.  All the calculations used the
Wilson gauge action.

Hart and Teper measured the glueball correlators, using only Wilson
loop interpolating operators, on this data set~\cite{Hart:2001fp}.
They found that the $0^{++}$ mass was significantly smaller in $n_f=2$
QCD, by a factor of ($\sim 0.84 \pm 0.03$), over quenched QCD at
comparable lattice spacing.  In figure~\ref{eq:GlueCont} we plot the
masses from the calculation by Hart and Teper (diamonds) with the
masses obtained in this analysis (bursts).  The inclusion of the
$\overline{\psi}\psi$ operators with the Wilson loop operators has
produced a further suppression of the mass of the $f_0$ at the lattice
spacing we use. The combined use of both Wilson loop and
$\overline{\psi}\psi$ operators is clearly a superior technique for
extracting masses than just using the Wilson loop operators on their
own, hence the result for the $f_0$ mass from this analysis is now the
preferred result.

Figure~\ref{eq:GlueCont} shows the suppression of the mass of the $f_0$
between $n_f=2$ and quenched QCD is less at this new lattice spacing
than our previous result at $a \sim 0.13 fm$~\cite{McNeile:2000xx}.
To make physical predictions
about the spectrum of the $f_0$ hadrons requires the continuum 
limit to be taken for the $n_f=2$ results.
In
quenched QCD it was found necessary 
to use a lattice spacing of 0.05 fm~\cite{Bali:1993fb}, this
size of lattice spacing will also be required for full QCD
calculations that use the Wilson gauge action.  This will be
computationally expensive~\cite{Wittig:2002hk} with clover fermions.



\end{document}